\documentclass[aps,prd,onecolumn,showpacs,amsmath,amssymb,nofootinbib]{revtex4}
\usepackage{amsfonts}
\usepackage{amsmath}
\usepackage{graphicx}
\usepackage{subfigure}
\usepackage{dcolumn}
\usepackage{bm}
\usepackage{booktabs}
\usepackage[utf8]{inputenc}
\usepackage{multirow}
\usepackage{graphicx,graphics,dcolumn,booktabs,bm}
\usepackage{longtable,lscape}
\usepackage{txfonts}
\usepackage{overpic}
\usepackage{amssymb}
\usepackage{indentfirst}
\usepackage{epsfig}
\usepackage{feynmf}   %{feynmp}
\usepackage{epstopdf}   %{feynmp}
\usepackage{slashed}  %for Feynman symbols
\usepackage{color}
\usepackage{ulem}

\def\f(s){[(\alpha+\beta)m_c^2-\alpha\beta s]}
\def\non{\\ \nonumber}
\def\bea{\begin{eqnarray}}
\def\eea{\end{eqnarray}}
\def\bq{\begin{quote}}
\def\eq{\end{quote}}

\def\nnb{\nonumber}

\def\rar{\rightarrow}														
\def\nnb{\nonumber}
\def\la{\langle}
\def\ra{\rangle}

\def\gg2{ \la\alpha_s G^2 \ra}
\def\gg3{g^3f_{abc}\la G^aG^bG^c \ra}

\begin{document}

\title{Investigation of the light four-quark states with exotic $J^{PC}=0^{--}$}
\author{Zhuo-Ran Huang$^{1,2}$}
\author{Wei Chen$^2$}
\email{wec053@mail.usask.ca}
\author{T. G. Steele$^2$}
\email{tom.steele@usask.ca}
\author{Zhu-Feng Zhang$^{2,3}$}
\email{zhangzhufeng@nbu.edu.cn}
\author{Hong-Ying Jin$^1$}
\email{jinhongying@zju.edu.cn}

\affiliation{
$^1$Zhejiang Institute of Modern Physics, Zhejiang University, Zhejiang Province, 310027, P. R. China
\\
$^2$Department of Physics and Engineering Physics, University of Saskatchewan, Saskatoon,
Saskatchewan, S7N 5E2, Canada
\\
$^3$Physics Department, Ningbo University, Zhejiang Province, 315211, P. R. China
}

\begin{abstract}
We study the exotic $J^{PC}=0^{--}$ four-quark states in Laplace sum rules (LSR) and finite energy sum rules (FESR). We use the vector tetraquark-like currents as interpolating currents in the correlator, from which the $1^{+-}$ states are also studied. In the mass extraction, we use the standard stability criterion with respect to the Borel parameters and the QCD continuum thresholds and consider the effect of the violation of factorization in estimating the high dimensional condensates as a source of uncertainties. The obtained mass prediction $1.66\pm0.14$ GeV is much lower than the previous sum rule predictions obtained using the scalar currents. Our result favours the four-quark interpretation of the possible $\rho\pi$ dominance in the $D^0$ decay. We also discuss the possible decay patterns of these exotic four-quark states.
\end{abstract}

\maketitle

%================================================================================
%================================================================================
\section{Introduction}\label{sec:Introduction}
%================================================================================
%================================================================================
The Dalitz-plot distribution of the $D^0\to\pi^+\pi^-\pi^0$ process was analyzed by the Babar Collaboration
\cite{2007-Aubert-p251801-251801,2008-Gaspero-p14015-14015}. The resonant dominance sub-structure
of this Cabibbo suppressed decay was then studied and indicated that an isospin-zero final state may exist
\cite{2010-Gaspero-p242-246,2010-Bhattacharya-p96008-96008,2015-Gronau-p114018-114018}. The analysis
of the Dalitz-plot behavior showed the typical structure of a $\pi^+\pi^-\pi^0$ final state with $I^GJ^{PC}=0^-0^{--}$
\cite{1964-Zemach-p1201-1201}, which is exotic and
cannot be composed of a quark-antiquark pair in the conventional quark model \cite{2014-Olive-p90001-90001,2007-Klempt-p1-202}.
If such a resonance exists near $M(D^0)\simeq1865$ MeV, it might be a hybrid or four-quark state \cite{2015-Gronau-p114018-114018}.

A hybrid meson is composed of a quark-antiquark pair and an excited gluonic field. It provides a good platform
to search for exotic quantum numbers that cannot be realized for a $q\bar q$ state. The mass of a $0^{--}$
hybrid was predicted to be about 1.8--2.2 GeV in the constituent gluon model \cite{1993-Ishida-p179-198},
around $2.3$ GeV in the QCD Coulomb gauge approach \cite{2007-General-p347-358} and 2.8--3.3 GeV in QCD sum rules \cite{Chetyrkin:2000tj,Govaerts:1984bk}.
These values and ranges are all higher than or marginally consistent with the mass of $D^0$ meson.
In the MIT bag model
\cite{1983-Barnes-p241-241,1983-Chanowitz-p211-211}, the hybrid states in the lightest supermultiplet with
quantum numbers $J^{PC}=(0, 1, 2)^{-+}, 1^{--}$ consist of a S-wave color-octet quark-antiquark pair
coupled to an excited gluonic field with $J_g^{P_gC_g}=1^{+-}$. A higher supermultiplet contains hybrids with
$J^{PC}=0^{+-}$, $(1^{+-})^3$, $(2^{+-})^2$, $3^{+-}$, $(0, 1, 2)^{++}$, which were composed of a P-wave
$q\bar q$ pair and the same gluonic excitation \cite{2012-Liu-p126-126, 2011-Dudek-p74023-74023}.
The hybrid state with $J^{PC}=0^{--}$ may lie higher than other channels and couple to a different gluonic
excitation. Such supermultiplet structures were confirmed in Lattice QCD \cite{2012-Liu-p126-126}, QCD
sum rules \cite{2013-Chen-p19-19,2014-Chen-p25003-25003} and the P-wave quasigluon approach
\cite{2008-Guo-p56003-56003}.

In quantum field theory, a hybrid $\bar qgq$ operator and a four-quark operator can transform into each other with
the same quantum numbers via quark annihilation interactions ($q\bar q\to g\to q\bar q$) in the $I_{q\bar q}=0$
channel. They tend to mix and couple to the same physical state. In general,
there are two types of four-quark operators: tetraquark-like operators $(qq)(\bar q\bar q)$ and molecule-like operators $(q\bar q)(q\bar q)$. They
are related to each other by the Fierz transformation and color rearrangement \cite{2007-Chen-p94025-94025}. The tetraquark formalism was first suggested
in the bag model by Jaffe \cite{1977-Jaffe-p267-267,1977-Jaffe-p281-281}, then it was extensively investigated
and used to study the nature of exotic hadron states \cite{2010-Chen-p105018-105018,2011-Chen-p34010-34010,
2016-Chen-p1-121,1980-Aerts-p1370-1370,2004-Maiani-p212002-212002,2005-Maiani-p14028-14028}.  In
the QCD Coulomb gauge approach, the masses of the $0^{--}$ molecules and tetraquarks were predicted to be
around 1.36 GeV and 2.15 GeV respectively in Refs.~\cite{2007-Cotanch-p656-661,2007-General-p216-223}.

The four-quark states with $J^{PC}=0^{--}$ were also systematically studied in the approach
of QCD sum rules using the scalar interpolating currents in Ref.~\cite{2009-Jiao-p114034-114034}, which seems not to support a mass below 2 GeV.
However, because the coefficient of the four-quark condensate is zero \cite{2009-Jiao-p114034-114034}\footnote{We have confirmed this result through independent calculation.}, it is difficult to assess the uncertainties associated with truncation of the OPE.
Therefore it is worth considering different interpolating currents in the sum rule analysis, which provide a chance to obtain OPE series that have better behaviours. In this work, we shall use the vector currents which can couple to $J^{PC}=0^{--}$ and $1^{+-}$ four-quark states. We shall perform the numerical analyses of the mass for both the scalar and vector channels. In order to make robust estimates, we shall use Laplace sum rules (LSR) and finite energy sum rules (FESR), and use the standard stability criterion with respect to the Borel parameter $\tau$ and the continuum threshold $s_0$ to extract the masses. We shall conclude the paper by discussing the possible explanation of the $\rho\pi$ dominance in $D^0$ decay and the decay patterns of the $0^{--}$ four-quark states.

%================================================================================
\section{Laplace sum rules and finite energy sum rules}\label{curr}
%================================================================================
%================================================================================

Introduced by SVZ in 1979 \cite{Shifman:1978bx}, QCD sum rules has become a powerful method to study the
hadronic properties. The basic idea of this approach is to relate the QCD expression of the correlation function (obtained using the well-known operator product
expansion) with the phenomenological parametrization by using the standard dispersion relation. The two-point correlation function of the vector current has the
following Lorentz structure:
\bea
\Pi_{\mu\nu}(q^{2}) & = & i\int d^{4}xe^{iqx}\left\langle 0\left|T\left[j_{\mu}(x)j_{\nu}^{+}(0)\right]\right|0\right\rangle \label{eq:1}\\
 & = & (q_{\mu}q_{\nu}-q^{2}g_{\mu\nu})\Pi_{v}(q^{2})+q_{\mu}q_{\nu}\Pi_{s}(q^{2})\nonumber,
\eea
where $j_{\mu}(x)$ in this work can be the four-quark currents that couple to both the $1^{+-}$ and $0^{--}$ states, and
the invariants $\Pi_{v}(q^{2})$ and $\Pi_{s}(q^{2})$ correspond
respectively to contributions from the $1^{+-}$ and $0^{--}$ states.

The correlation function obeys the dispersion relation, which relates the $\Pi(q^2)$ with its imaginary part $\textrm{Im}\Pi(q^2)$.
For hadrons with light flavour quarks, the dispersion relation reads:
\begin{eqnarray}
\Pi_{v/s}(q^{2})=\frac{1}{\pi}\int_{0}^{\infty}ds\frac{\textrm{Im}\Pi_{v/s}(s)}{s-q^{2}-i\epsilon}.
\label{eq:20}
\end{eqnarray}

The dispersion relation provides an important connection between QCD and phenomenology as on the theoretical side the correlator can be
expanded in terms of QCD vacuum condensates for large Euclidean momentum (i.e. $Q^2=-q^2$ is much greater than the QCD scale $\Lambda_{\rm QCD}$) while on the
phenomenological side the spectral function at low energy can be measured  and parameterized experimentally. In this work, we adopt the widely used ``one single narrow resonance minimal duality ansatz" (which has been tested in $e^+e^-\rightarrow \rm hadrons$ and charmonium data \cite{Narison:2002pw,Narison:2002hk,Narison:1980ti}) to parametrize the spectral function as:
\bea
\nonumber
\frac{1}{\pi}\textrm{Im}\Pi_{v/s}(s)&\simeq&\sum_n\delta(s-m_n^2)\langle0|\eta|n\rangle\langle n|\eta^+|0\rangle\\
&\simeq&f_H^2\delta(s-m_H^2)+ \mbox{``QCD continuum"}\times\theta(s-s_0),
\label{res_model}
\eea
where $f_H$ and $s_0$ respectively denote the the coupling of the current to the hadron and the QCD continuum threshold.

On the QCD side, the correlation function can be calculated perturbatively using the Operator Product Expansion:
\bea
\Pi_{v/s}(q^2)
\simeq \sum_{d=0,2,4,...}\frac{1}{( q^2 )^{d/2}}
\sum_{dim O=d} C(q^2)~\langle 0|O|0\rangle~,
\eea
where $\langle 0|O|0\rangle~$ are the QCD vacuum condensates of dimension $d$,  $C(q^2)$  are the corresponding wilson coefficients calculated in the perturbation theory. For the large Euclidian $q^2$, the OPE reaches good convergence thus the first few terms (up to condensate dimension 8 in this work) are expected to be a good approximation of the correlation function. To further improve the convergence of OPE and also to suppress the
continuum contribution in the spectral integral, one can apply the inverse Laplace operator to both sides of the dispersion relation and then the moment and the ratio of  Laplace/Borel sum rules (LSR) can be derived:
\bea\label{usr}
{M}_{v/s}(\tau,s_0)
&=& \int_{0}^{s_0} {ds}~\mbox{exp}(-s\tau)
~\frac{1}{\pi}~\mbox{Im} \Pi_{v/s}(s)~,\\{R}_{v/s}(\tau,s_0)
&=& -\frac{d}{d\tau} \log {
M_{v/s}}\simeq M_H^2~.
\eea

Due to the uncertainties induced by the truncation of the OPE series and the simple parametrization of the spectral function, the output results depend on the two external parameters $(\tau,s_0)$. SVZ originally suggested the sum rule should be analyzed within a certain range of the Borel parameter $\tau$, which ensures
both the validity of the OPE truncation and the suppression of the continuum contribution to the spectral integral. Furthermore, some attempts to determine $(\tau,s_0)$
objectively from the sum rules have also been made following either the standard stability criterion with respect to  $(\tau,s_0)$ \cite{Narison:2002pw,Narison:2009vj,Matheus:2004qq,Matheus:2006xi}\footnote{For recent examples using the stability criterion see \cite{Albuquerque:2016znh} for traditional QCD sum rules and \cite{Huang:2016upt} for light-cone QCD sum rules.} or the original concept of the sum rule window \cite{Shifman:1978bx} (e.g., the Monte-carlo based weighted-least-squared matching
procedure \cite{Leinweber:1995fn,Zhang:2013rya,Huang:2014hya}\footnote{The Holder inequalities can provide constraints on the sum rule window as discussed in \cite{Benmerrouche:1995qa,Benmerrouche:1995pi,Steele:1996np}
}).
One would expect to optimize the output results by demanding
that they are insensitive to the variation of $(\tau,s_0)$. However, as is well-known, LSR for multi-quark currents (of high dimension) is more likely than those for ordinary $q\bar{q}$ measons to suffer from  Eq.~\eqref{res_model}, the simple
parametrization of the spectral function, which could lead to the absence of the $s_0$-stability. Such cases have occurred in tetraquark and pentaquark sum rules, as has been discussed in \cite{Narison:2009vj,Matheus:2004qq,Matheus:2006xi}.

In the case where the $s_0$-stability is not reached, FESR has been shown in some other sum rule analyses for multi-quark states \cite{Narison:2009vj,Matheus:2004qq,Matheus:2006xi} to be a useful complement. The moment and ratio of FESR read:
\bea\label{fesr}
{\cal R}_{v/s}(s_0)
&=&{\int_{0}^{s_0} {ds}~s
~\frac{1}{\pi}~\mbox{Im} \Pi_{v/s}(s)\over
\int_{0}^{s_0} {ds}
~\frac{1}{\pi}~\mbox{Im} \Pi_{v/s}(s)} \simeq M_H^2~,
\eea
which provides a connection between the lowest state mass and the continuum threshold. FESR can be obtained by letting the Borel parameter $\tau$ be zero in LSR prior to renormalization-group improvement, thus it's quite natural to expect such an approach can help reduce the effects of high dimensional condensates in the sum rules and provide the possibility to restore the $s_0$-stability.

\section{Interpolating currents for $0^{--}$/$1^{+-}$ light four-quark states}
The $0^{--}$ light four-quark states have been studied in \cite{2009-Jiao-p114034-114034} using the scalar diquark-antidiquark currents, which does not support a mass below 2 GeV. However, as noted earlier, the coefficients of four-quark condensates have been found to be zero in the OPE \cite{2009-Jiao-p114034-114034}, which raises some doubts on the resulting accuracy of the sum rules. Furthermore, the $s_0$-stability is not reached in \cite{2009-Jiao-p114034-114034}, suggesting that the currents used in \cite{2009-Jiao-p114034-114034} may not provide sufficiently reliable sum rules. Therefore, here we use diquark-antidiquark vector currents which can couple to both the $1^{+-}$ and $0^{--}$ four-quark states.

The Lorentz structures of the $1^{+-}/0^{--}$ diquark anti-diquark vector currents have been systematically studied in \cite{2011-Chen-p34010-34010} for the charmonium-like states. Here we use  the $ud\bar{u}\bar{d}$ currents of the same Lorentz structures (under isospin symmetry, $uu\bar{u}\bar{u}$  and $dd\bar{d}\bar{d}$ share the same sum rules with $ud\bar{u}\bar{d}$ at leading order):
\begin{eqnarray}
\nonumber J_{1\mu}&=&u^T_aCd_b(\bar{u}_a\gamma_{\mu}\gamma_5C\bar{d}^T_b+\bar{u}_b\gamma_{\mu}\gamma_5C\bar{d}^T_a)
-
u^T_aC\gamma_{\mu}\gamma_5d_b(\bar{u}_aC\bar{d}^T_b+\bar{u}_bC\bar{d}^T_a)\,
, \non J_{2\mu}&=&u^T_aCd_b(\bar{u}_a\gamma_{\mu}\gamma_5C\bar{d}^T_b-\bar{u}_b\gamma_{\mu}\gamma_5C\bar{d}^T_a)
-
u^T_aC\gamma_{\mu}\gamma_5d_b(\bar{u}_aC\bar{d}^T_b-\bar{u}_bC\bar{d}^T_a)\,
, \non J_{3\mu}&=&u^T_aC\gamma_5d_b(\bar{u}_a\gamma_{\mu}C\bar{d}^T_b+\bar{u}_b\gamma_{\mu}C\bar{d}^T_a)
-
u^T_aC\gamma_{\mu}d_b(\bar{u}_a\gamma_5C\bar{d}^T_b+\bar{u}_b\gamma_5C\bar{d}^T_a)\,
,
\\ \label{currents2}
J_{4\mu}&=&u^T_aC\gamma_5d_b(\bar{u}_a\gamma_{\mu}C\bar{d}^T_b-\bar{u}_b\gamma_{\mu}C\bar{d}^T_a)
-
u^T_aC\gamma_{\mu}d_b(\bar{u}_a\gamma_5C\bar{d}^T_b-\bar{u}_b\gamma_5C\bar{d}^T_a)\,
, \non J_{5\mu}&=&u^T_aC\gamma^{\nu}d_b(\bar{u}_a\sigma_{\mu\nu}\gamma_5C\bar{d}^T_b+\bar{u}_b\sigma_{\mu\nu}\gamma_5C\bar{d}^T_a)
-
u^T_aC\sigma_{\mu\nu}\gamma_5d_b(\bar{u}_a\gamma^{\nu}C\bar{d}^T_b+\bar{u}_b\gamma^{\nu}C\bar{d}^T_a)\,
, \non J_{6\mu}&=&u^T_aC\gamma^{\nu}d_b(\bar{u}_a\sigma_{\mu\nu}\gamma_5C\bar{d}^T_b-\bar{u}_b\sigma_{\mu\nu}\gamma_5C\bar{d}^T_a)
-
u^T_aC\sigma_{\mu\nu}\gamma_5d_b(\bar{u}_a\gamma^{\nu}C\bar{d}^T_b-\bar{u}_b\gamma^{\nu}C\bar{d}^T_a)\,
, \non J_{7\mu}&=&u^T_aC\gamma^{\nu}\gamma_5d_b(\bar{u}_a\sigma_{\mu\nu}C\bar{d}^T_b+\bar{u}_b\sigma_{\mu\nu}C\bar{d}^T_a)
-
u^T_aC\sigma_{\mu\nu}d_b(\bar{u}_a\gamma^{\nu}\gamma_5C\bar{d}^T_b+\bar{u}_b\gamma^{\nu}\gamma_5C\bar{d}^T_a)\,
, \non J_{8\mu}&=&u^T_aC\gamma^{\nu}\gamma_5d_b(\bar{u}_a\sigma_{\mu\nu}C\bar{d}^T_b-\bar{u}_b\sigma_{\mu\nu}C\bar{d}^T_a)
-
u^T_aC\sigma_{\mu\nu}d_b(\bar{u}_a\gamma^{\nu}\gamma_5C\bar{d}^T_b-\bar{u}_b\gamma^{\nu}\gamma_5C\bar{d}^T_a)\,
,
\end{eqnarray}
where $J_{1\mu}$, $J_{3\mu}$, $J_{5\mu}$, $J_{7\mu}$ have the color structure $\mathbf 6 \otimes \mathbf {
\bar 6}$ and $J_{2\mu}$, $J_{4\mu}$, $J_{6\mu}$, $J_{8\mu}$ have the color structure $\mathbf { \bar 3 }\otimes \mathbf 3$. Since the states of different isospin are degenerate in masses at leading order (LO), we do not differentiate the isospin in our calculation.

\section{QCD expressions for the two-point correlation functions}

After performing the SVZ expansion in the chiral limit ($m_u=m_d=0$) to LO
of the perturbation series, we arrive at the following expression for the correlation function (up to dimension-8 condensate contributions) resulting from $J_i$ ($i=1-8$):
\begin{equation}
\label{eq:xx}
\frac{1}{\pi}\mbox{Im}\Pi_{i;s/v}(s) = a_{i;s/v} \frac{s^3}{\pi^6}+ b_{i;s/v} \frac{\langle\alpha_s G^2\rangle s}{\pi^5} + c_{i;s/v} \frac{\langle \bar qq\rangle^2}{\pi^2}+ d_{i;s/v} \frac{\langle \bar qGq\rangle \langle \bar qq\rangle}{\pi^2 s},
\end{equation}
where the coefficients $a_{i;v/s} - d_{i;v/s}$ are listed in Table~\ref{tab:coeff}.\footnote{Here we omit the dimension-6 and dimension-8 gluon condensate contributions which are suppressed by a loop factor. The complete evaluation of the dimension-8 quark and gluon condensate contributions considering operator mixing under renormalization was done in \cite{Huang:2014hya} for the $1^{-+}$ light hybrid meson, where the contributions from the gluon condensates are comparable to those from the quark condensates.}

\begin{table}[htbp]
\centering
\begin{tabular}{|c|c|c|c|c|c|c|c|c|}
\hline
\multirow{2}{*}{}& \multicolumn{8}{c|}{$i$}
  \\
  \cline{2-9}
   & 1 & 2 & 3 & 4 & 5 & 6 & 7 & 8 \\
  \hline
  \hline
 $a_{i;s}$ & 1/30720 &  1/61440 &1/30720 & 1/ 61440 & 1/10240 & 1/20480 & 1/ 10240 & 1/20480 \\
  \hline
  $b_{i;s}$ & -1/1536 & 1/1536 &-1/1536 & 1/1536 & 11/1536 & 1/1536 & 11/1536 & 1/1536 \\
  \hline
  $c_{i;s}$ & 1/6  &1/12  &-1/6  & -1/12 & -5/6 & -5/12 & 1/6 & 1/12 \\
  \hline
  $d_{i;s}$ &1/8 &1/16  & -1/8 & -1/16 & -5/8 & -5/16 & 1/8 & 1/16 \\
  \hline
     \hline
  $a_{i;v}$ & 1/18432 &  1/36864 & 1/18432 & 1/36864 & 1/ 6144 & 1/12288 & 1/ 6144&1/ 12288 \\
  \hline
  $b_{i;v}$ & -1/4608 & 1/4608 & -1/4608 & 1/4608 & 11/ 4608 &1/4608 & 11/4608& 1/ 4608 \\
  \hline
  $c_{i;v}$ & -5/18 & -5/36 & 5/18 & 5/36 & 25/18 & 25/36 & -5/18& -5/36 \\
  \hline
  $d_{i;v}$ & -1/8 &  -1/16 & 1/8 & 1/16 & 5/8 & 5/16 & -1/8 & -1/16 \\
  \hline
\end{tabular}
\caption{\label{tab:coeff} The coefficients for Eq.\eqref{eq:xx}.}
\end{table}

However, a vacuum-factorization violation factor has been noticed for a long time in the process $e^+ e^-\rightarrow hadrons$ \cite{Bertlmann:1984ih,Bertlmann:1987ty,Launer:1983ib,Narison:1992ru,Narison:1995jr}
and $\tau$ decay \cite{Narison:2009vy}. Therefore it's necessary to consider the errors induced by the violation of factorization in our numerical analysis.
For the condensates $d\leq6$ and the QCD scale which are under good control from the experiments, we shall use the values given in Table \ref{param_tab}.

\begin{table}
\begin{tabular}{ll}
  \hline
   & Reference \\
   \hline
  $\la\alpha_s G^2\ra \simeq (7\pm2)\times10^{-2}~\rm{GeV}^4$ & sum rules of $ e^+e^-\rar \rm{hadrons}$ \cite{Launer:1983ib,Narison:1995jr,Eidelman:1978xy}
and J/$\Psi$ \cite{Narison:2011xe,Narison:2011rn} \\
  $g\la\bar{\psi}G\psi\ra\equiv g\la\bar{\psi}\frac{\lambda_a}{2}\sigma^{\mu\nu}G^a_{\mu\nu}\psi\ra\simeq
(0.8\pm
0.1)~{\rm GeV}^2\la\bar \psi\psi\ra$ & light baryon systems \cite{Dosch:1988vv,Ioffe:1981kw} \\
  $\rho\alpha_s  \la\bar \psi\psi\ra^2\simeq
(4.5\pm 0.3) \times 10^{-4}~\rm{ GeV}^6$\footnotemark[4] & $e^+e^-\rar \rm{hadrons}$ \cite{Narison:1992ru,Narison:1995jr} and $\tau$-decay \cite{Narison:2009vy} \\
  $\Lambda_{\rm QCD}= (353\pm 15)~{\rm MeV}$ & $\tau$-decay \cite{Narison:2009vy} \\
  \hline
\end{tabular}
\caption{QCD parameters used in our analysis. The quantity $\rho$ indicates the violation of factorization hypothesis in estimating the four-quark condensates.}
\label{param_tab}
\end{table}

%\footnotetext[4]{\tadd{The quantity} $\rho$ indicates the violation of factorization in estimating the four-quark condensates.}
%%%%% where $\rho$ indicates the violation of factorization; ======================================

\subsection{masses of the $0^{--}$ four-quark states}
As mentioned previously, the sum rules for currents of different color structures at LO are only slightly different  (we will see this more obviously in this and the next subsection). What affects the behaviour of the sum rules considerably  are the Lorentz structures of the currents, by which we will categorize the currents in our analyses. For each category, if stability is reached (with a suitably converging OPE for LSR), we will extract the mass considering the effects of violation of factorization as a source of theoretical uncertainty in our analysis.

%%%%%%%%%%%%%%%%%%%%%%%%%%%%%%%1====ok
\begin{figure}[htbp]
\centering
\subfigure[]{
\includegraphics[scale=0.69]{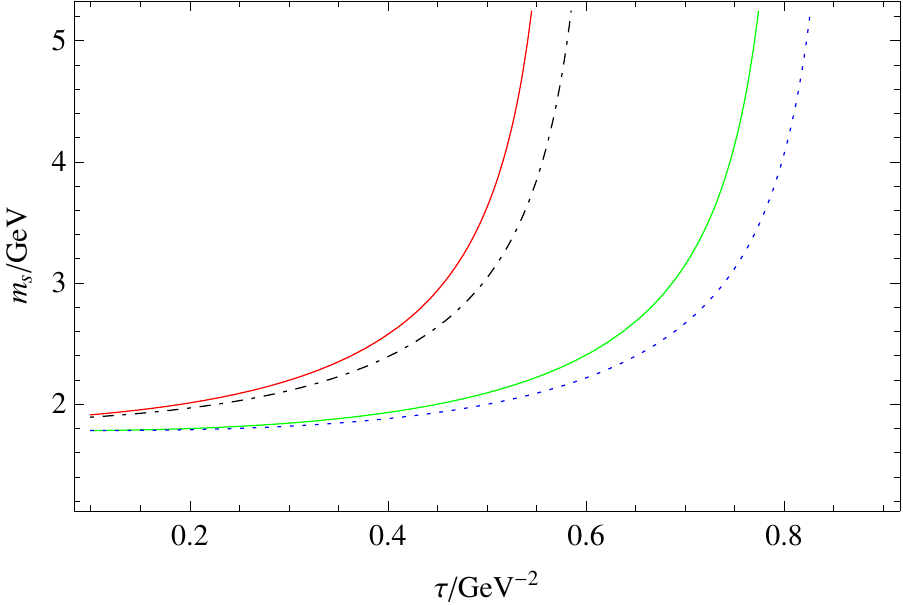}}
\subfigure[]{
\includegraphics[scale=0.7]{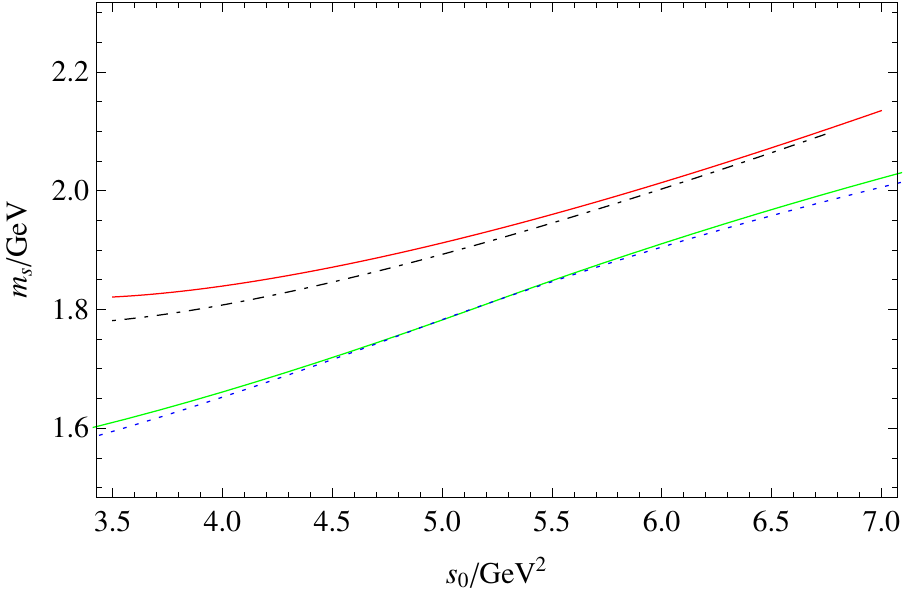}}
\caption{\label{fig:slj1j2}(a) The $0^{--}$ four-quark masses versus
$\tau$ obtained from the LSR for $J_{1\mu}$ (green continuous), from the LSR for $J_{1\mu}$ with violation of factorization by a factor $\rho=2$ in estimating the dimension-8 condensates (red continuous), from the LSR for $J_{2\mu}$ (blue dotted) and from the LSR for $J_{2\mu}$ with violation of factorization by a factor $\rho=2$ in estimating the dimension-8 condensates (black dotted-dashed); (b) the same as (a) but for the masses versus $s_0$.
}
\end{figure}
%%%%%%%%%%%%%%%%%%%%%%%%%%%%%%%%%%%%

\begin{figure}[htbp]
\centering
\includegraphics[scale=0.7]{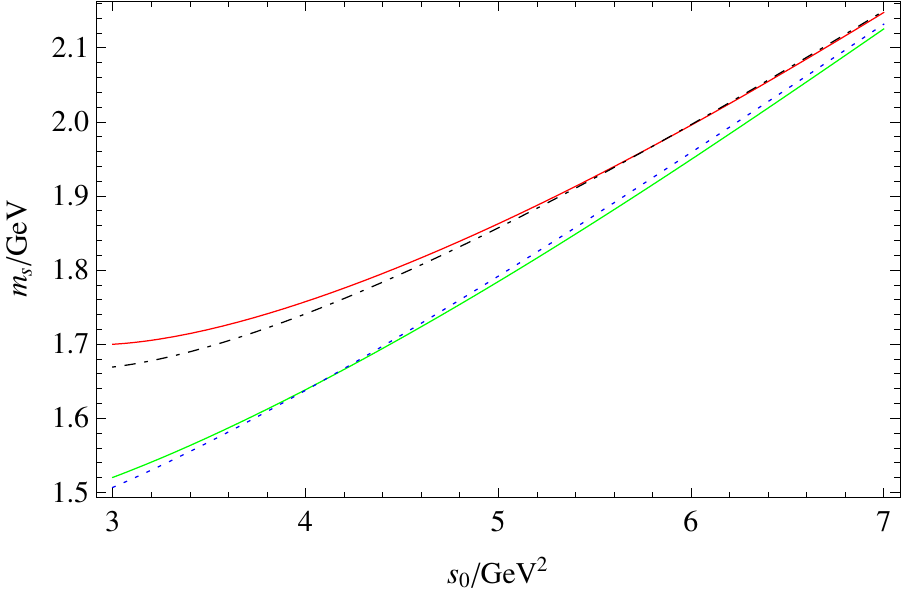}
\caption{\label{fig:sfj1j2}
The $0^{--}$ four-quark masses versus $s_0$
obtained from the FESR for $J_{1\mu}$ (green continuous), from the FESR for $J_{1\mu}$ with violation of factorization by a factor $\rho=2$ in estimating the dimension-8 condensates (red continuous), from the FESR for $J_{2\mu}$ (blue dotted) and from the FESR for $J_{2\mu}$ with violation of factorization by a factor of $\rho=2$ in estimating the dimension-8 condensates (black dotted-dashed).
}
\end{figure}

We first consider the LSR for $J_{1\mu}$ and $J_{2\mu}$. As shown in FIG. \ref{fig:slj1j2}, the LSR for both $J_{1}$ and $J_2$ reach the $\tau$-stability (which is lost after considering violation of factorization). However, the masses obtained from the $\tau$-stability points increase gradually with $s_0$, which means the $s_0$-stability is not reached and thus $s_0$ cannot be determined from the LSR for $J_{1}$ and $J_2$. We consider the following values obtained at $s_0=5.0 {\rm GeV}^2$ (assuming the mass of the lowest-lying state is below 2 GeV, which is consistent with the subsequent analyses resulting from other currents):
\bea
M_{s1;L}&=&1.78~{\rm GeV}~at~s_0=5.0~{\rm GeV^2},\nnb\\
M_{s2;L}&=&1.78~{\rm GeV}~at~s_0=5.0~{\rm GeV^2}.
\eea
The FESR for $J_1$ and $J_2$ also do not reach stability, which can be seen in FIG. \ref{fig:sfj1j2}. Therefore no results can be obtained from the FESR.

%%%%%%%%%%%%%%%%%%%%%%%%%%%%%%%%%%%%2-------------OK
%%%%%%%%%%%%%%%%%%%%%%%%%%%%%%%%%%%%%%%%%

$J_3$--$J_6$ belong to two different Lorentz structures, but these four currents have similar sum rule behaviour, thus we present their results together.
%%%%%%%%%%%%%%%%%%%%%3=====OK
\begin{figure}[htbp]
\centering
\subfigure[]{
\includegraphics[scale=0.7]{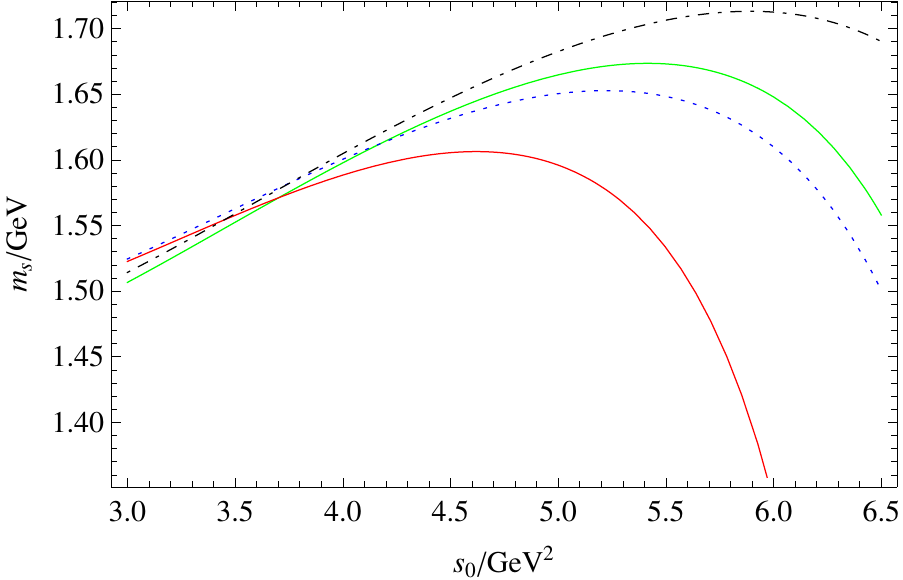}}
\subfigure[]{
\includegraphics[scale=0.7]{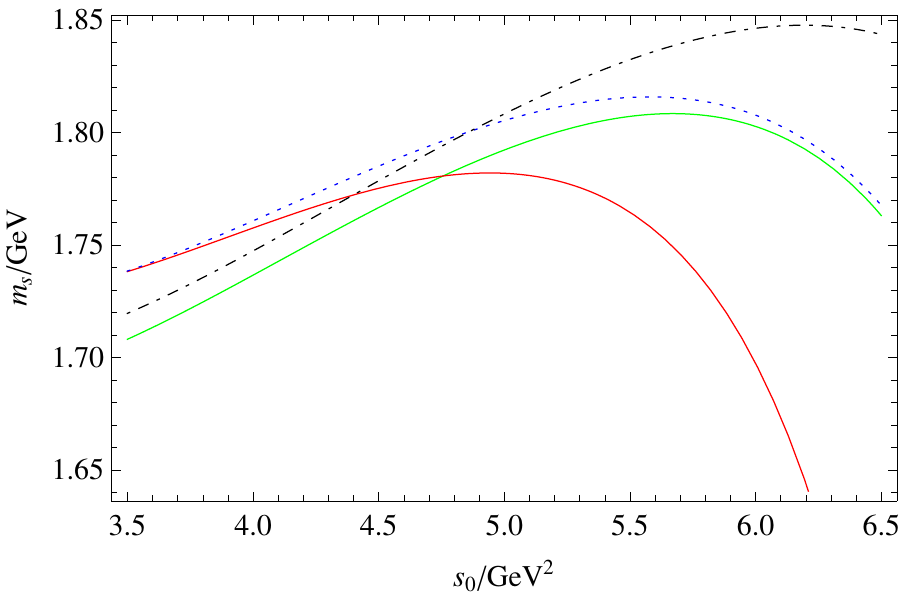}}
\caption{\label{fig:sfj3tj6} (a) The $0^{--}$ four-quark masses versus $s_0$
obtained from the FESR for $J_{3\mu}$ (green continuous), from the FESR for $J_{4\mu}$ (red continuous), from the FESR for $J_{5\mu}$ (blue dotted) and from the
FESR for $J_{6\mu}$ (black dotted-dashed); (b) the same as (a) but for masses versus $s_0$ from the FESR with violation of factorization by a factor $\rho=2$ in estimating the dimension-8 condensates.
}
\end{figure}
%%%%%%%%%%%%%%%%%%%%%
Contrary to $J_1$ and $J_2$, $J_3$--$J_6$ have worse LSR but better FESR behaviour.
The LSR ratios for $J_{3\mu}$--$J_{6\mu}$ monotonically increase with the Borel parameter $\tau$ thus no ($\tau,s_0$) stability is reached in LSR, while the FESR ratios reach stability in $s_0$ as shown in FIG \ref{fig:sfj3tj6}, which allow the following mass predictions:
\bea
M_{s3;F}&=&1.67(1.81)~{\rm GeV}~at~s_0=5.4(5.7) ~{\rm GeV^2},\nnb\\
M_{s4;F}&=&1.61(1.73)~{\rm GeV}~at~s_0=4.6(5.0)~ {\rm GeV^2},\nnb\\
M_{s5;F}&=&1.65(1.82)~{\rm GeV}~at~s_0=5.2(5.6) ~{\rm GeV^2},\nnb\\
M_{s6;F}&=&1.71(1.85)~{\rm GeV}~at~s_0=5.9(6.2) ~{\rm GeV^2},\nnb
\eea
where we have also presented the mass predictions (in the brackets) obtained considering the deviation of the dimension-8 condensates from their factorized values.

For $J_7$  and $J_8$, the LSR moment ratios begin to reach the $\tau$-stability at $s_0=3.8$ and $s_0=4.2$ respectively. But again the $s_{0}$-stability
is absent. If we use $s_0=5~ \rm GeV^2$ (as we did for $J_1$ and $J_2$, which can be justified by the FESR for $J_3$--$J_6$), we obtain

\bea
M_{s7;L}&=&1.80(1.95)~{\rm GeV}~at~s_0=5.0(5.5)~{\rm GeV^2},\nnb\\
M_{s8;L}&=&1.83(1.97)~{\rm GeV}~at~s_0=5.0(5.5)~{\rm GeV^2}.
\eea
Unfortunately, the FESR ratios for both $J_7$ and $J_8$ do not reach stability (plots are not shown here for simplicity). On the contrary, they increase with $s_0$, which means no results can be obtained from these sum rules.

In our analysis of the scalar channel, the FESR ratios have better behaviour than the LSR ratios which do not reach the $s_0$--stability for all the currents. Therefore we consider the results from FESR as more reliable. From the FESR for $J_{3\mu}$--$J_{6\mu}$ where stability is reached, we fix the mass of the $0^{--}$ four-quark state to be $M_s=1.66\pm0.14$ GeV,\footnote{The errors come from the discrepancies between the results obtained from using different currents, the violation of factorization hypothesis in estimating dimension-8 quark condensates and the errors of the QCD input parameters.} which is justified by the LSR results obtained from using the $s_0$ deduced from FESR given that the central values obtained from LSR are consistent with the FESR results within the errors.
%%%%%%%%%%%%%%%%%%%%%%%%%%%%%%%%%4?? without label?????????????????????????????????????-------------OK
\begin{figure}[htbp]
\centering
\includegraphics[scale=0.7]{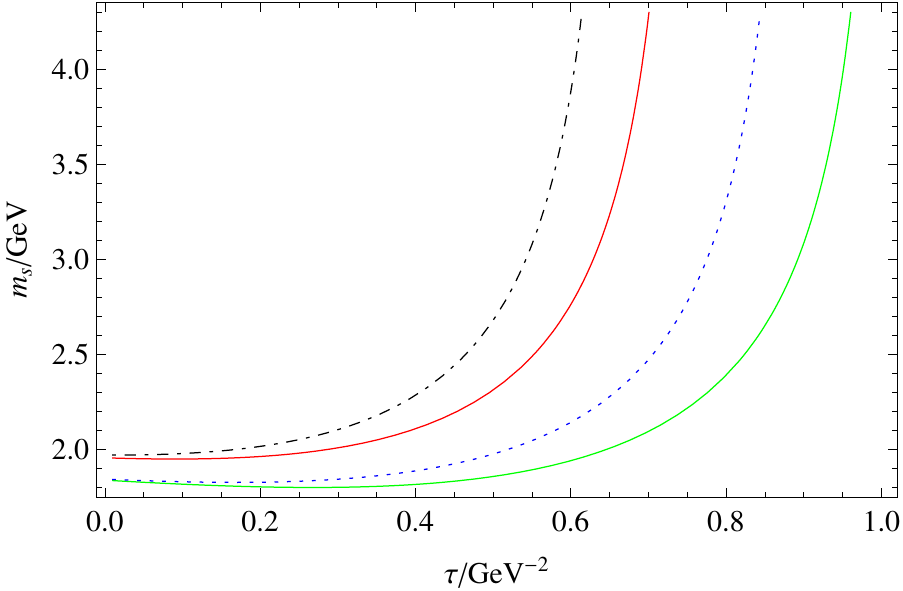}
\caption{\label{fig:slj7j8}The $0^{--}$ four-quark masses versus $\tau$
obtained from the LSR for $J_{7\mu}$ (green continuous), from the LSR for $J_{7\mu}$ with violation of factorization by a factor $\rho=2$ in estimating the dimension-8 condensates (red continuous), from the LSR for $J_{8\mu}$ (blue dotted) and from the LSR for $J_{8\mu}$ with violation of factorization by a factor $\rho=2$ in estimating the dimension-8 condensates (black dotted-dashed).
}
\end{figure}
%%%%%%%%%%%%%%%%%%%%%%%%%%%%%%

\subsection{masses of the $1^{+-}$ four-quark states}

%%%%%%%%%%%%%%%%%%%%%%%%%%5======OK
\begin{figure}[htbp]
\centering
\subfigure[]{
\includegraphics[scale=0.7]{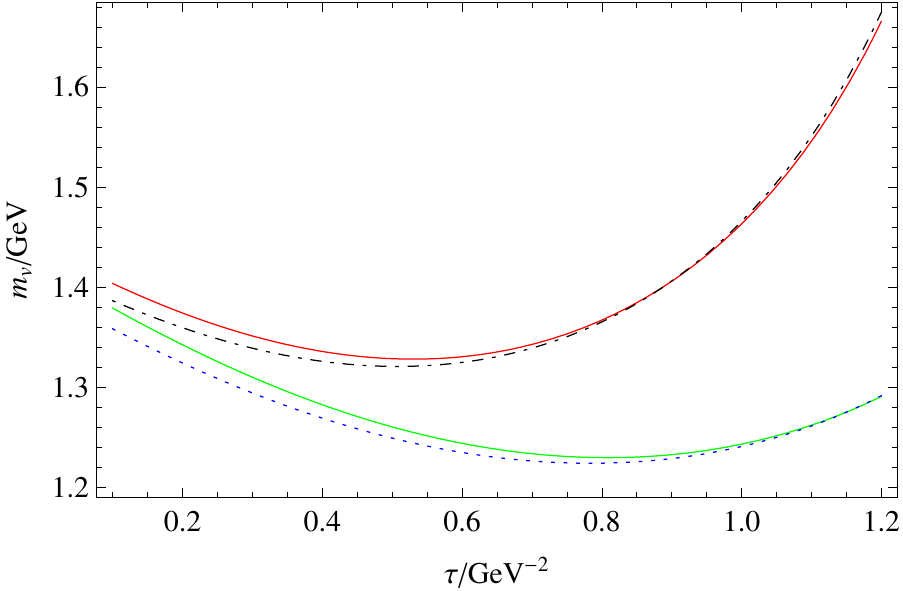}}
\subfigure[]{
\includegraphics[scale=0.7]{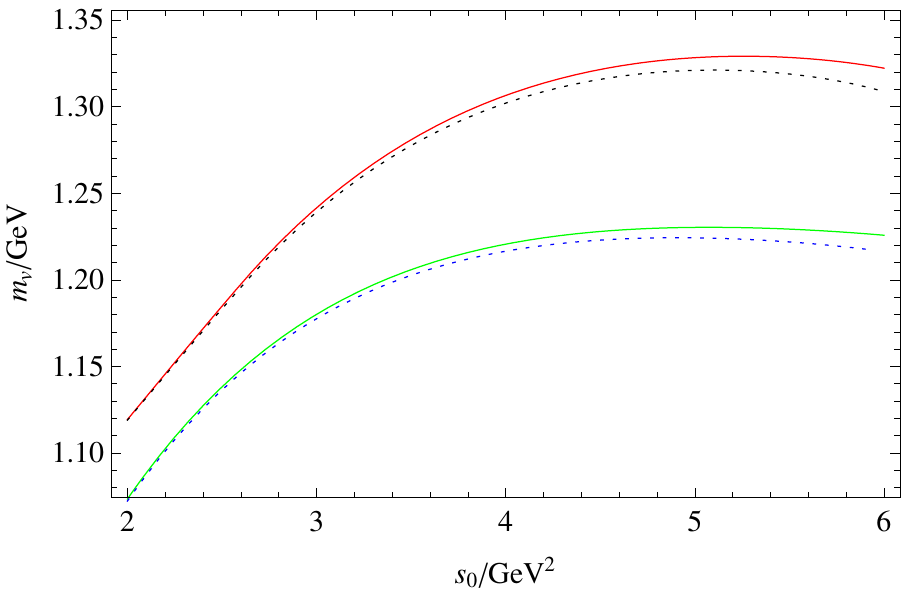}}
\caption{\label{fig:vlj1j2}
(a) The $1^{+-}$ four-quark masses versus
$\tau$ obtained from the LSR for $J_{1\mu}$ (green continuous), from the LSR for $J_{1\mu}$ with violation of factorization by a factor $\rho=2$ in estimating the dimension-8 condensates (red continuous), from the LSR for $J_{2\mu}$ (blue dotted) and from the LSR for $J_{2\mu}$ with violation of factorization by a factor $\rho=2$ in estimating the dimension-8 condensates (black dotted-dashed); (b) The same as (a) but for masses versus $s_0$.
}
\end{figure}
%%%%%%%%%%%%%%%%%%%%%%%%%%%%%%%

Similarly, we use the stability criterion for extracting the mass of the $1^{+-}$ four-quark states.
For $J_{1\mu}$ and $J_{2\mu}$, LSR reach both the $\tau$ and $s_0$ stability as shown in FIG. \ref{fig:vlj1j2}, from which we obtain the
following predictions:
\bea
M_{v1;L}&=&1.23(1.33)~{\rm GeV}~at~s_0=4.8(5.0)~ {\rm GeV^2},\nnb\\
M_{v2;L}&=&1.22(1.32)~{\rm GeV}~at~s_0=4.8(5.0)~ {\rm GeV^2},
\eea
where we have presented the results (in the brackets) with violation of factorization, and we have also checked that dimension-8 condensate contributions constitute less than 18$\%$ of the
total QCD expression, which ensures the validity of the truncation of OPE.
We also use the FESR as a reexamination of the results obtained from the LSR. For $J_{1\mu}$ and $J_{2\mu}$, the following results can
be obtained from the stability points of the FESR ratios:
\bea
M_{v1;F}&=&1.42(1.44)~{\rm GeV}~at~s_0=5.1(5.3) ~{\rm GeV^2},\nnb\\
M_{v2;F}&=&1.40(1.42)~{\rm GeV}~at~s_0=4.9(5.1)~ {\rm GeV^2}.
\eea

%%%%%%%%%%%%%%%%%%%%%%%%%%%%%%6?????????????without label???????????????--------------OK
\begin{figure}[htbp]
\centering
\includegraphics[scale=0.7]{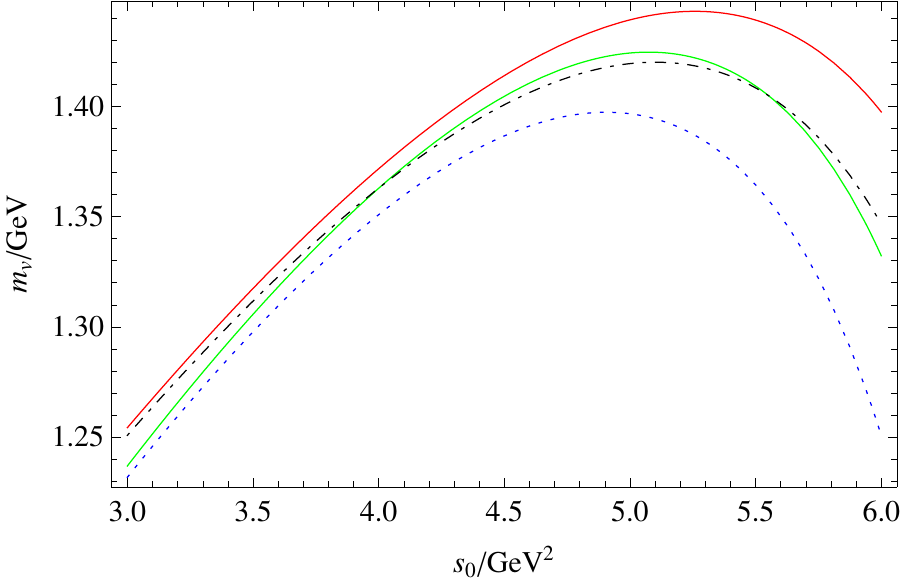}
\caption{The $1^{+-}$ four-quark masses versus $s_0$
obtained from the FESR for $J_{1\mu}$ (green continuous), from the FESR for $J_{1\mu}$ with violation of factorization by a factor $\rho=2$ in estimating the dimension-8 condensates (red continuous), from the FESR for $J_{2\mu}$ (blue dotted) and from the FESR for $J_{2\mu}$ with $\rho=2$ violation of factorization by a factor of 2 in estimating the dimension-8 condensates (black dotted-dashed).
}
\end{figure}
%%%%%%%%%%%%%%%%%%%%%%%%%%%%%%%

One can see that the LSR results differ from the FESR results by 180--190 MeV here, which may result from the non-inclusion of the radiative corrections and as
well the simple ``one resonance + continuum" parametrization for the spectral
function. Therefore, it's appropriate to consider (in the final analysis) a conservative range of the mass according to the predictions obtained from using different sum rules.

As in the scalar channel, $J_{3\mu}$--$J_{6\mu}$ (belonging to two different lorentz structures) also have quite similar sum rules. The LSR ratios show $\tau$-stability  but no $s_0$-stability. Using the $s_0$ fixed from the LSR and FESR for $J_{1\mu}$ and $J_{2\mu}$, all these moments give a mass around 1.27 $\rm GeV$. We shall not consider this value in our final determination of the mass due to the absence of the $s_0$-stability. The FESR for $J_{3\mu}$--$J_{6\mu}$ do not reach stability in  $s_0$ either, which means no reliable predictions can be obtained
from the sum rules for $J_{3\mu}$--$J_{6\mu}$.

\begin{figure}[htbp]
\centering
\subfigure[]{
\includegraphics[scale=0.7]{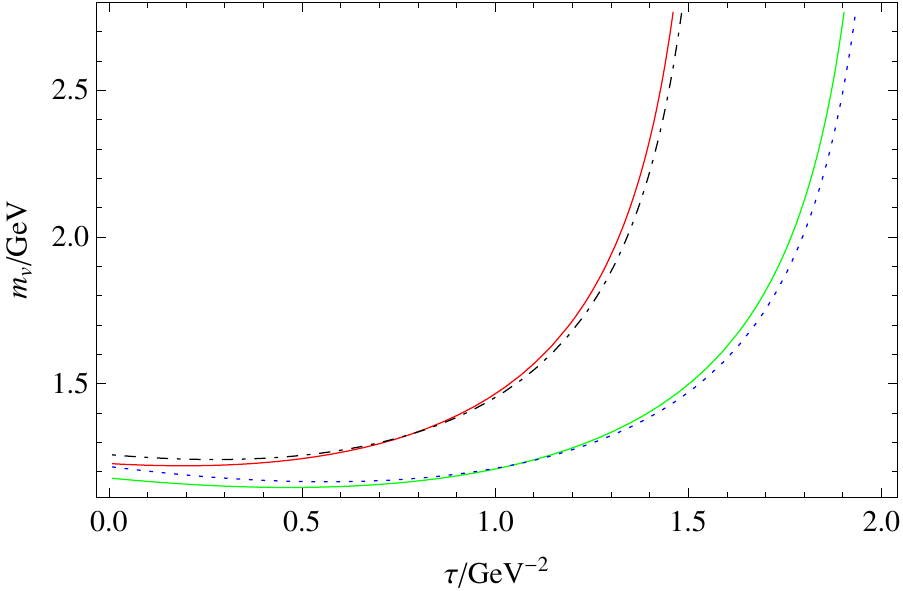}}
\subfigure[]{
\includegraphics[scale=0.7]{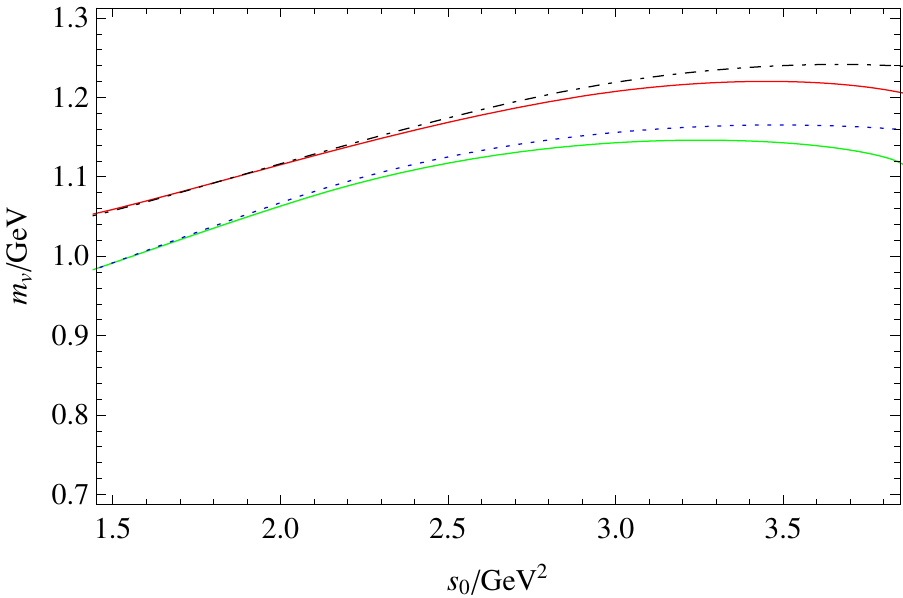}}
\caption{\label{fig:vlj7j8}
(a) The $1^{+-}$ four-quark masses versus
$\tau$ obtained from the LSR for $J_{7\mu}$ (green continuous), from the LSR for $J_{7\mu}$ with violation of factorization by a factor $\rho=2$ in estimating the dimension-8 condensates (red continuous), from the LSR for $J_{8\mu}$ (blue dotted) and from the LSR for $J_{8\mu}$ with violation of factorization by a factor $\rho=2$ in estimating the dimension-8 condensates (black dotted-dashed); (b) the same as (a) but for masses versus $s_0$.
}
\end{figure}
%%%%%%%%%%%%%%%%%%

For $J_{7\mu}$ and $J_{8 \mu}$, using the stability criterion for  ($\tau$,$s_0$), we obtain from the LSR ratios (see FIG. \ref{fig:vlj7j8}) the following optimal values:
\bea
M_{v7;L}&=&1.15(1.22)~{\rm GeV}~at~s_0=3.2(3.4) ~{\rm GeV^2},\nnb\\
M_{v8;L}&=&1.17(1.24)~{\rm GeV}~at~s_0=3.5(3.6) ~{\rm GeV^2}.
\eea

Although the LSR ratios reach the ($\tau$,$s_0$) stability, we will not use these results in our final estimate due to the dimension-8 condensate contributions constituting more than half of the total OPE expressions (due to the cancellation between lower dimensional condensate
terms and the perturbative terms), which makes the truncation of OPE unreliable.

The FESR for $J_{7\mu}$ and $J_{8 \mu}$ also reach stability, which occur at lower $s_0$ compared with the sum rules for $J_{1\mu}$ and $J_{2\mu}$. However, the
predicted masses are comparable to those obtained using other currents, which read:
\bea
M_{v7;F}&=&1.18(1.23)~{\rm GeV}~at~s_0=3.3(3.5) ~{\rm GeV^2},\nnb\\
M_{v8;F}&=&1.22(1.26)~{\rm GeV}~at~s_0=3.5(3.7) ~{\rm GeV^2}.
\eea

%%%%%%%%%%%%%%8????????????????without label?????????????????????????????????????????????======OK
\begin{figure}[htbp]
\centering
\includegraphics[scale=0.7]{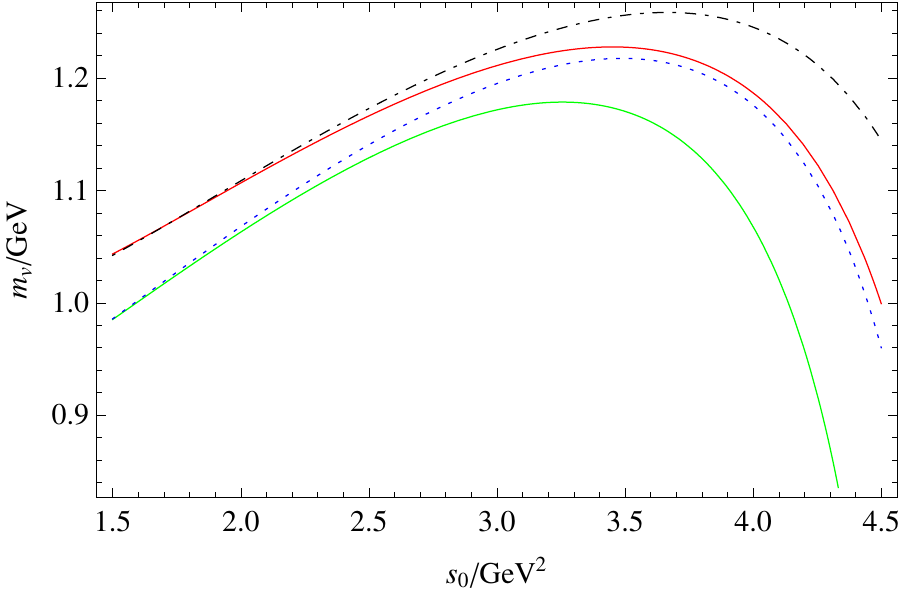}
\caption{The $1^{+-}$ four-quark masses versus $\tau$
obtained from the FESR for $J_{7\mu}$ (green continuous), from the FESR for $J_{7\mu}$ with violation of factorization by a factor $\rho=2$ in estimating the dimension-8 condensates (red continuous), from the FESR for $J_{8\mu}$ (blue dotted) and from the FESR for $J_{8\mu}$ with violation of factorization by a factor $\rho=2$ in estimating the dimension-8 condensates (black dotted-dashed).
}
\end{figure}
%%%%%%%%%%%%%

From the above results we can see that the LSR for $J_{1\mu}$ and $J_{2\mu}$ and the FESR for $J_{7\mu}$ and $J_{8\mu}$ reach stability. From these sum rules, we can obtain the optimal result for the $1^{+-}$ four-quark state. Considering the discrepancy between different sum rules, we shall consider a conservative range instead of extracting a central value. The obtained mass range is $M_{v}$=1.18--1.43 GeV.

\section{Discussion and conclusions}
We have studied the $0^{--}$ and $1^{+-}$ light four-quark states using Laplace and finite energy sum rules. In the scalar channel, LSR and FESR give consistent mass predictions (within the errors), which we have determined to be $1.66\pm0.14$ GeV. The non-negligible errors mainly come from the violation of factorization in estimating the dimension-8 condensates. Our predictions for $0^{--}$ four-quark states are significantly lower than the ones obtained in \cite{2009-Jiao-p114034-114034} using different interpolating currents, where a mass below 2 GeV is not supported but the results suffer from the absence of four-quark condensates in the OPE. In contrast to the previous work, the results in this work do not exclude the possibility of the $\rho\pi$ dominance in $D^0$ decay to be a four-quark state. Given that the $D^0$ mass is much lower than the mass prediction for a $0^{--}$ hybrid state in the QCD Coulomb gauge approach \cite{2007-General-p347-358} and QCD sum rules \cite{Chetyrkin:2000tj} and barely covered by the range predicted in the constituent gluon models \cite{Ishida:1991mx,General:2006ed}, the four-quark explanation seems to be more reasonable. Moreover, the mixing of the four-quark state and hybrid state is another possible explanation that needs to be considered, which we hope to discuss in the future.

Following our prediction, we can discuss the decay patterns of the $0^{--}$ four-quark states. Considering the kinematical constraints and
the conservation of $I, G, J, P, C$ we find the following two-body hadronic decay modes:
\begin{equation}
X_{0^-0^{--}}\to \rho\pi, \, \omega\eta, \, f_0h_1;~X_{1^+0^{--}}\to a_0\pi,\, \omega\pi, \, \rho\eta;~X_{2^-0^{--}}\to \rho\pi.
\end{equation}

The $\rho\pi$ decay mode of the isoscalar state can be the observed channel (via $\rho\to\pi^+\pi^-$) in Babar \cite{2007-Aubert-p251801-251801}. If this $3\pi$ resonance exists, it may also be seen in the $\omega\eta$ final states.
The isospin partner states are expected to be observed in the above final states, among which the charged $a_0\pi$ final states are worth special attention, for they are
the only possible S-wave decay mode, the others are in P-wave.

For the vector channel, there exist differences of approximately 200 MeV  between the LSR and FESR results, which can result from the simple parametrization on the phenomenological side and the lack of radiative corrections in the OPE. We have conservatively estimated the mass to be in the range 1.18--1.43 GeV, which suggest the $1^{+-}$ four-quark states lie within the 260 MeV range above the conventional $q\bar{q}$ state $h_1(1170)$.

\begin{acknowledgments}
This work is supported by NSFC under grant 11175153, 11205093 and 11347020, and supported by K. C. Wong Magna Fund in Ningbo University. TGS is supported by the Natural Sciences and Engineering Research Council of Canada (NSERC). Z.R. Huang and Z.F. Zhang are grateful to the University of Saskatchewan for its warm hospitality.
\end{acknowledgments}

%\bibliography{myreference}

\end{document}